\documentclass[superscriptaddress,aps,prl,twocolumn]{revtex4}
\usepackage{bm}
\usepackage{ulem}
\usepackage{epsfig}
\usepackage{graphicx}
\usepackage{amssymb,amsmath,amsbsy,amsgen,amsfonts}    
\usepackage{dcolumn}
\usepackage{amsthm}
\usepackage{mathrsfs}
\usepackage{latexsym}
\usepackage{array}
\usepackage{color}
\usepackage{amstext}
\allowdisplaybreaks[1]
\usepackage{txfonts}
\usepackage{pifont}
\usepackage{epstopdf} 

\newcommand{\bra}[1]{\left\langle{#1}\right\vert}
\newcommand{\ket}[1]{\left\vert{#1}\right\rangle}

\newcommand{\be}{\begin{equation}}
\newcommand{\ee}{\end{equation}}
\newcommand{\ba}{\begin{array}}
\newcommand{\ea}{\end{array}}
\newcommand{\bqa}{\begin{eqnarray}}
\newcommand{\eqa}{\end{eqnarray}}

\DeclareSymbolFont{symbols}{OMS}{cmsy}{m}{n}

\begin{document}

\title{Active control of a plasmonic metamaterial for quantum state engineering}

\author{S. A. Uriri}
\affiliation{School of Chemistry and Physics, University of KwaZulu-Natal, Durban 4001, South Africa}
\author{T. Tashima}
\affiliation{School of Chemistry and Physics, University of KwaZulu-Natal, Durban 4001, South Africa}
\author{X. Zhang}
\affiliation{School of Chemistry and Physics, University of KwaZulu-Natal, Durban 4001, South Africa}
\author{M. Asano}
\affiliation{Department of Material Engineering Science, Graduate School of Engineering Science, Osaka University, Osaka 560-8531, Japan}
\author{M. Bechu}
\affiliation{Institute of Applied Physics, Karlsruhe Institute of Technology, 76128 Karlsruhe, Germany}
\affiliation{Institute of Nanotechnology, Karlsruhe Institute of Technology, 76128 Karlsruhe, Germany}
\author{\\ D. \"O. G\"uney}
\affiliation{Department\,of\,Electrical\,and\,Computer\,Engineering, Michigan Technological University, Houghton, MI 49931, USA}
\author{T. Yamamoto}
\affiliation{Department of Material Engineering Science, Graduate School of Engineering Science, Osaka University, Osaka 560-8531, Japan}
\author{\c{S}. K. \"Ozdemir}
\affiliation{ Department of Engineering Science and Mechanics, Pennsylvania State University, University Park, Pennsylvania 16802, USA}
\author{M. Wegener}
\affiliation{Institute of Applied Physics, Karlsruhe Institute of Technology, 76128 Karlsruhe, Germany}
\affiliation{Institute of Nanotechnology, Karlsruhe Institute of Technology, 76128 Karlsruhe, Germany}
\author{M. S. Tame}
\email{markstame@gmail.com}
\affiliation{School of Chemistry and Physics, University of KwaZulu-Natal, Durban 4001, South Africa}

\date{\today}

\begin{abstract}
We experimentally demonstrate the active control of a plasmonic metamaterial operating in the quantum regime. A two-dimensional metamaterial consisting of unit cells made from gold nanorods is investigated. Using an external laser we control the temperature of the metamaterial and carry out quantum process tomography on single-photon polarization-encoded qubits sent through, characterizing the metamaterial as a variable quantum channel. The overall polarization response can be tuned by up to 33\% for particular nanorod dimensions. To explain the results, we develop a theoretical model and find that the experimental results match the predicted behavior well. This work goes beyond the use of simple passive quantum plasmonic systems and shows that external control of plasmonic elements enables a flexible device that can be used for quantum state engineering.
\end{abstract}


\maketitle

{\it Introduction.---} Metamaterials are artificially engineered materials that provide optical, mechanical or thermal responses beyond what can be achieved by conventional materials~\cite{Lee12,Christ15,Cummer16,Cai10}. In optics, metamaterials have traditionally been made from metallic nanostructures much smaller than the wavelength of interest, with the collective behavior of many nanostructures giving rise to the bulk response of the material~\cite{Soukoulis11}. The use of metal in the design of optical metamaterials is mainly due to the ability to modify the electric and magnetic resonances of the nanostructures in the optical band easily using plasmonic techniques~\cite{Maier07}. However, the use of dielectric metamaterials has also been considered for achieving similar resonance behaviour~\cite{Jahani16}. In recent years, researchers have studied in great detail the electric and magnetic resonances of plasmonic nanostructures, using them to design metamaterials for a wide range of applications, including new types of lenses and imaging devices~\cite{Zhang08}, transformation optics components~\cite{Chen10}, sensing platforms~\cite{Chen12}, and many others~\cite{Billings13}. 

Most recently, studies have started to probe optical metamaterials in the quantum regime~\cite{Wang12,Zhou12,Chipouline12,Cortes14,Scholl15,McEnery14,Moiseev10,Siomau12,Zhou16,Tame13,Zago16}, with applications in quantum state engineering, such as entanglement distillation~\cite{Asano15,Faro15}, remote quantum interference~\cite{Jha15}, quantum state generation~\cite{Gheer16,Podd16} and dissipative-based quantum state control~\cite{Roger15,Altuzarra17}. However, a highly desirable behavior of metamaterials is the ability to control their response. Much work has been done in the classical regime in demonstrating metamaterials that can be actively controlled~\cite{Boardman11,Hess12,Zheludev12}. Techniques used include adding phase change materials~\cite{Gavinit72,Jepsen06,Prayakarao16,Wang15}, embedding liquid crystals~\cite{Werner07,Khoo10,Shrek13}, using mechanical deformation~\cite{Gutruf16, Ee16,Liu17,Kamali16}, electrical stimulation~\cite{Feig10,Zheludev12} and exploiting thermal effects~\cite{Driscoll08,Liu16,Lewan15}. In the quantum regime, there have not yet been any studies showing the active control of a metamaterial. Given the many quantum applications of metamaterials already demonstrated, it is important to investigate the possibility of active control and develop flexible components for quantum state engineering tasks.

In this work we experimentally demonstrate the active control of a metamaterial in the quantum regime. We investigate a two-dimensional metamaterial, a metasurface~\cite{Kildishev13,Yu14,Meinzer14,Xia14,Lin14}, that is comprised of unit cells made from gold nanorods, and whose usefulness was recently shown in an experiment performing entanglement distillation of partially entangled two-photon states~\cite{Asano15}. Here, we go beyond this work to show how the response of this metasurface can be actively controlled. Such control is useful in a real-world application in quantum communication~\cite{Gisin07}, where a single metasurface whose properties can be tuned to cover a wide operating range is desired in order to avoid the need to physically change the metasurface being used. To provide this tunability, we control the metasurface via its thermal response. We use an external laser to vary the temperature optically and carry out quantum process tomography on single photons sent through, with the goal of characterizing the metasurface as a variable quantum channel in the polarization basis. The process tomography method provides full information about how the metamaterial responds in the quantum regime under an external stimulus and enables us to assess the quality of its quantum operation. 

We find that with the correct nanorod dimensions one can achieve tuneability of the metasurface's polarization response by up to 33\%. To help explain the results, a theoretical model is developed and used. We find the experimental results match well the predicted behavior. Overall, the result is rather surprising given that we are increasing the amount of loss in the nanorods by heating -- one would expect the loss to have a negative impact on the operation of the metasurface. However, it is exactly the extra loss that changes the resonance properties of the nanorods and modifies the bulk response of the metasurface. Our work goes beyond previous studies using simple passive plasmonic systems in the quantum regime and shows that external control of plasmonic elements provides a versatile metamaterial that can be used to carry out quantum state engineering.
\begin{figure*}[t]
\centering
\includegraphics[width=17cm]{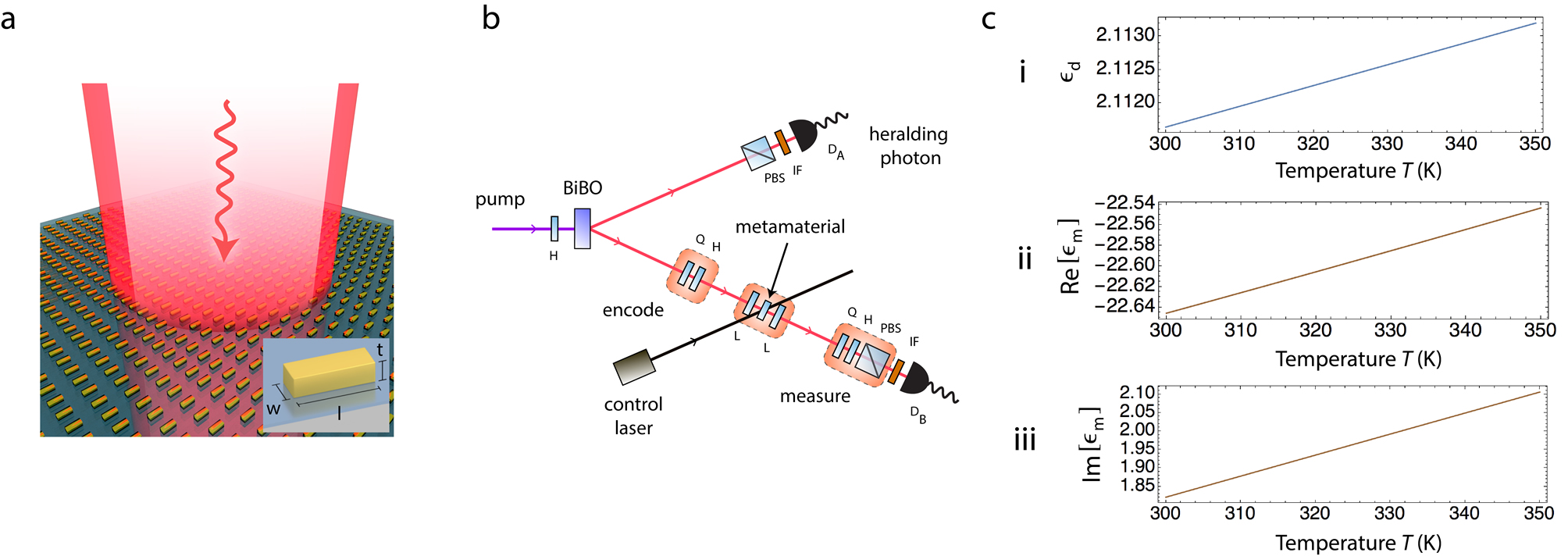}
\caption{\label{fig1} Overview and experimental setup for demonstrating active control of a metamaterial in the quantum regime. (a) Pictorial representation of one of the metamaterials used with a single photon (red) and an active control laser beam (white) sent through. The spot size of the control and single-photon beams are the same in the experiment, however the control beam is shown as smaller for pictorial purposes. The inset shows a 3D figure of the nanorods in each unit cell (dimensions considered are given in the main text). (b) The experimental setup, where a nonlinear BiBO crystal is pumped at 405~nm, producing pairs of photons at 810~nm via spontaneous parametric down-conversion. One photon is detected at detector $D_A$ and heralds the presence of a single photon in the other arm. Here, H is a half-wave plate, Q is a quarter-wave plate, L is a plano-convex lens ($f=25$~mm), PBS is a polarizing beamsplitter, IF is an interference filter ($\lambda=810$~nm and $\Delta \lambda=10$~nm), and $D_A$ and $D_B$ are single-photon detectors. (c) Temperature dependence of the permittivity, $\epsilon_d$ of the silica substrate (i), and that of the gold used for the nanorods, ${\rm Re}[\epsilon_m]$ (ii) and ${\rm Im}[\epsilon_m]$ (iii).}
\end{figure*}

{\it Overview.---} A diagram of the scenario for demonstrating the active control of a metamaterial in the quantum regime is shown in Fig.~\ref{fig1}~(a). Here, a single-photon (red beam) and an external control laser for heating (white beam) are incident on the metamaterial. The inset shows the geometry of the nanorods in each unit cell. In Fig.~\ref{fig1}~(b) the optical setup used is shown. Pairs of photons are generated via parametric down-conversion using a pump laser~\cite{Burnham70,Hong86}, with one photon of the pair acting as a heralding photon (top path) and the other as a probe photon (bottom path). Quantum information is encoded in the polarization degree of freedom of the probe single photons, with each photon representing a qubit. The qubits are encoded in different states and then sent through the metamaterial. The metamaterial is then characterized as a quantum channel as the temperature is changed via the control laser. Further details of the setup are given later in the experimental setup section.

{\it Theoretical model.---} 
Before introducing the experimental setup and analysing the results, we briefly summarize our theoretical model for the temperature dependence of a given metamaterial. The transmission response of the type of plasmonic metamaterial fabricated and shown in Fig.~\ref{fig1}~(a) can be modelled as a periodic array of nanoparticles in a planar rectangular lattice with periods $d_x$ and $d_y$. In the dipole approximation, each nanoparticle representing a unit cell of the metamaterial is modelled by a dipole with a polarizability tensor ${\boldsymbol \alpha}$, which relates the dipole moment ${\bf P}$ to the local electric field ${\bf E}_0$, {\it i.e.} ${\bf P}={\boldsymbol \alpha} {\bf E}_0$~\cite{Alu11,Zhao11,Zhao13,Bohren83}. The plasmonic nanoparticles fabricated are rod-like in shape and are well described as an ellipsoid with semi-axes $a$, $b$ and $c$. This gives a diagonal polarizability tensor with non-zero elements~\cite{Bohren83}
\begin{equation}
\alpha_i = 4 \pi \epsilon_0 a b c \frac{\epsilon_{m}-\epsilon_d}{3\epsilon_d + 3 L_i(\epsilon_{m}-\epsilon_d)},
\end{equation}
where $\epsilon_0$ is the free-space permittivity, $L_i$ $(i = x,y,z$) is the geometrical factor~\cite{Bohren83}, $\epsilon_{m}$ is the relative permittivity of gold and $\epsilon_d$ is the relative permittivity of the surrounding medium (silica). To relate the semi-axes to the geometry of our nanorods we set $a=w/2$, $b=t/2$ and $c=l/2$, where $w$, $t$, and $l$ are the nanorod width, thickness and length, respectively, as shown in the inset of Fig.~\ref{fig1}~(a). 

The reflection and transmission through a given metamaterial nanorod array can be related to the polarizability tensor of the nanorod unit cell and the interaction dyadic of the array, as described in detail in Ref.~\cite{Alu11}. For light with normal incidence to the array and polarized in direction $k$ we have
 \begin{equation}
 T_k = 1+ \frac{i \mu_0 \pi f c}{d_x d_y}\frac{\alpha_k}{1 - \beta_k \alpha_k}
 \label{Trans}
 \end{equation}
and $R_k = T_k-1$, where $\mu_0$ is the free-space permeability, $f$ is the frequency of the propagating electromagnetic wave, $c$ is the speed of light in vacuum and $\beta_k$ is the interaction dyadic. Here, we set $\beta_k = 0$ due to the large nanorod spacing considered in our experiment. Note that in the special case of no absorption in the array the relation $|R|^2+|T|^2=1$ is satisfied. The final step is to introduce the temperature dependence of the permittivities $\epsilon_d$ and $\epsilon_m$, after which we are in a position to model transmission through the metamaterial of qubits encoded into single photons as horizontal and vertical polarization states. We choose the vertical axis of the single photons to be oriented along the long (length) axis of the nanorods.
\begin{figure}[t]
\centering
\includegraphics[width=8.3cm]{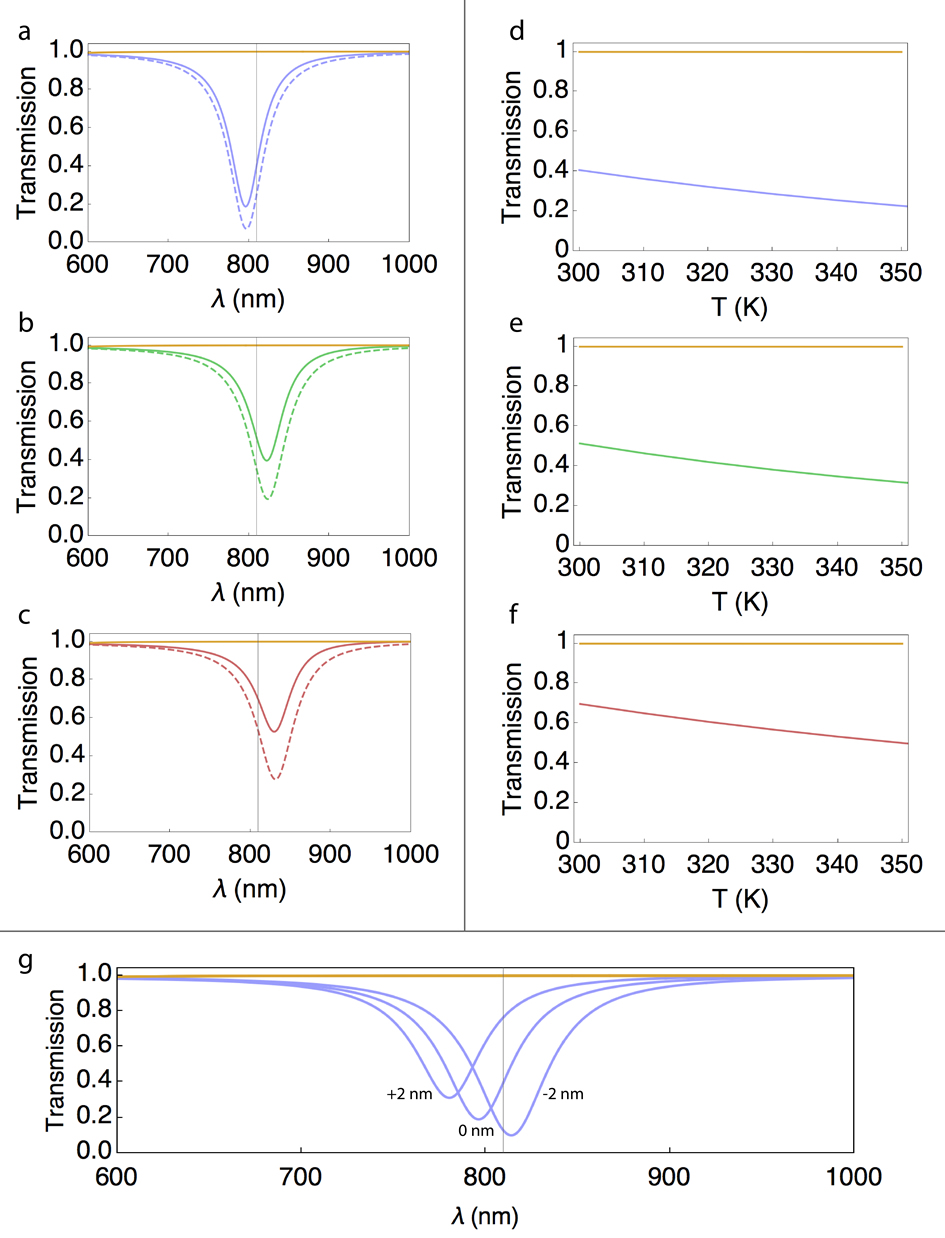}
\caption{\label{fig2} Temperature-dependent transmission response of metamaterials with different nanorod dimensions (theory). In panels (a)-(f), the period is fixed at $d_x = d_y = 200$~nm and the thickness is $t = 30$~nm. In panels (a)-(c), the lower solid resonance curve is for vertical transmission at $T=300$~K and the lower dotted resonance curve is for vertical transmission at $T=340$~K. The horizontal solid line is for horizontal transmission. The dimensions of the nanorods used are: (a) width $w=46$~nm and length $l = 130$~nm, (b) $w=47$~nm and $l = 140$~nm, and (c) $w=48$~nm and $l = 144$~nm. Panels (d)-(f) show the corresponding temperature dependence over a range of $50$~K at $\lambda=810$~nm with the nanorod dimensions chosen as those used in panels (a)-(c), respectively. (g) Transmission response of a metamaterial with nanorod dimensions corresponding to panel (a) in the middle, and with $+/- 2$~nm added to the length, width and thickness.}
\end{figure}

To model the temperature dependence of the silica substrate we use the wavelength-dependent thermo-optic coefficient $\frac{{\rm d} n}{{\rm d} T}$, where $n$ is the refractive index ($\epsilon_d=n^2$) and $T$ is the temperature~\cite{Lin07,Leviton08}. The refractive index is related to the thermo-optic coefficient by the relation 
\be
n(T) = n (T_r) + (T - T_r) \frac{{\rm d} n}{{\rm d} T},
\label{Dmodel}
\ee 
where $n(T)$ is the temperature dependent refractive index, which is also wavelength dependent, and $T_r=300~K$ is a reference temperature. It is known that the refractive index of fused silica at $T_r$ can be well described by the Sellmeier equation $n (T_r,\lambda) = [1 + A_1 \lambda^2/(\lambda^2 - \beta^2_1) + A_2 \lambda^2/(\lambda^2 - \beta^2_2) + A_3 \lambda^2/(\lambda^2 - \beta^2_3)]^{1/2}$, where the coefficients are $A_1 = 0.6961663$, $A_2 = 0.4079426$, $A_3 = 0.8974794$, $B_1 = 0.0684043$, $B_2 = 0.1162414$, $B_3 = 9.896161$ and $\lambda = 0.81~\mu$m is the wavelength of interest~\cite{Leviton08}. We then have $\frac{{\rm d} n}{{\rm d} T} = (GR + HR^2)/2 n (T_r,\lambda)$, with $R = \lambda^2/(\lambda^2 - \lambda^2_{ig})$, $\lambda_{ig}$ = 0.109 $\mu$m is the band-gap wavelength of silica, $G = -1.6548 \times 10^{-6}$ K$^{-1}$ and $H = 31.7794 \times 10^{-6}$  K$^{-1}$~\cite{Lin07}. The temperature dependence of $\epsilon_d$ is shown in Fig.~\ref{fig1}~(c)~(i) for $T=300$ to 350~K.

For the gold nanorods, the temperature dependence is described using a modified Drude model, valid below the interband transition frequency~2.4~eV ($\lambda \gtrsim 520$~nm)~\cite{Rakic98,Ozdemir03,Bouillard12}   
\begin{equation}
\epsilon_m (T) = \epsilon_ \infty - \frac{\omega_p^2(T)}{\omega (\omega + i \omega_c(T))},
\label{Mmodel}
\end{equation}
where $\omega$ is the angular frequency of the electromagnetic field, $\epsilon_ \infty$ is the high-frequency permittivity of the metal, and $\omega_p(T)$ and $\omega_c(T)$ represent the temperature-dependent plasmon frequency and collision frequency of the free electrons, respectively. The plasmon frequency is given by $\omega_p(T) = \omega_p(T_r)/[1 + 3\gamma(T-T_r)]^{1/2}$, where $\omega_p(T_r)$ is the plasmon frequency at the reference temperature and $\gamma$ = 14.2 $\times$ 10$^{-6}$ K$^{-1}$ is the thermal linear expansion coefficient. The collision frequency results from a combination of electron-electron and electron-phonon scattering, with $\omega_c(T)$ = $\omega_{e-e}(T)$ + $\omega_{e-ph}(T)$, where $\omega_{e-e}(T) = \pi^3 \Gamma \Delta[(K_B T)^2 + (\hbar \omega / 2 \pi)^2]/12 \hbar E_F$ and $\omega_{e-ph}(T) = \omega_0[2/5 + 4 (T/\theta_D)^5\int_0^{\frac{\theta_D}{T}}z^4(e^z - 1)^{-1}dz]$. Here, $k_B$ is the Boltzmann constant, $\hbar$ is Plank's constant,
$\theta_D$ is the Debye temperature, $E_F$ is the Fermi-level energy for gold, $\Gamma$ is the Fermi-surface average of scattering probability, $\Delta$ is the fractional Umklapp scattering coefficient and $\omega_0$ is a constant. The following parameters are used for the above equations: $\theta_D = 185$~K, $E_F = 5.5$~eV, $\Gamma = 0.55$ and $\Delta= 0.77$~\cite{Bouillard12}. Furthermore, the following parameters are obtained by fitting the experimental data for gold from Ref.~\cite{Rakic98} at the reference temperature ($\lambda \gtrsim 600$~nm) to Eq.~(\ref{Mmodel}): $\omega_0 = 0.346$~eV, $\epsilon_\infty = 8$ and $\omega_p(T_r)=53.41$~eV. The temperature dependence of $\epsilon_m$ is shown in Fig.~\ref{fig1}~(c)~(ii) for ${\rm Re}[\epsilon_m]$ and (iii) for ${\rm Im}[\epsilon_m]$, for $T=300$ to 350~K.

Using Eqs.~(\ref{Dmodel}) and (\ref{Mmodel}) in Eq.~(\ref{Trans}) we are able to model the temperature-dependent response of the metamaterial transmission. As an example, in Fig.~\ref{fig2}~(a) we show the transmission $|T|^2$ for horizontal and vertical polarized light over the wavelength range $600$-$1000$~nm for a metamaterial at two different temperatures ($T=300$~K and $T=340$~K). The dimensions used for the simulation are chosen based on the size of the nanorods fabricated (see experimental section) and given by $d_x = d_y = 200$~nm for the period, $t = 30$~nm for the thickness, $w=46$~nm for the width and $l = 130$~nm for the length. One can clearly see the change in the transmission for vertically polarized light as the temperature changes (lower solid and dotted curves), whereas the transmission for horizontally polarized light is not affected significantly (upper solid line). This contrast is due to the dependence of the vertical transmission coefficient on the plasmon resonance along the length of the nanorod, which is relatively strong and can change significantly depending on the value of the permittivity of the metal. On the other hand, for horizontally polarized light, the plasmonic resonance is weak along the width of the rod and so changes in the permittivity do not have a significant effect. In Figs.~\ref{fig2}~(a)-(c) and (g), a vertical line marks the wavelength of interest for our experiment ($\lambda=810$~nm). In Fig.~\ref{fig2}~(d) we show the temperature dependence of the transmission for $\lambda=810$~nm over the range $300$-$350$~K. In order to understand further how the transmission changes depending on the nanorod dimensions we show two more examples of metamaterials in Fig.~\ref{fig2}~(b), (c), (e) and (f). The dimensions used are the same as the previous example, but with $w=47$~nm and $l=140$~nm for Fig.~\ref{fig2}~(b) and (e), and $w=48$~nm and $l=144$~nm for Fig.~\ref{fig2}~(c) and (f). One can see that depending on the nanorod dimensions the value of the transmission for vertically polarized light can vary significantly as the temperature is modified.

In Fig.~\ref{fig2}~(g) we show how deviations in the nanorod dimensions ($\pm 2$~nm for $w$, $t$ and $l$) affect the transmission of vertically polarized light through the metamaterial at a fixed temperature of $300$~K. One can see that with only a small deviation of $2$~nm the transmission curve plotted as a function of the wavelength of the incident light is shifted considerably to the left (+2~nm) or to the right (-2~nm). This provides useful information about how a realistic metamaterial might respond, as consistency of nanorod dimensions across the array is hard to achieve during fabrication. Based on the behavior seen in Fig.~\ref{fig2}~(g), the result of this would be a linewidth broadening and a shift of the wavelength where the transmission becomes minimum (the resonance wavelength).

{\it Experimental setup.---} In Fig.~\ref{fig1}~(b) the optical setup used in our experiment is shown. Here, single photons are generated using Type-1 spontaneous parametric down-conversion. Pairs of single photons at $\lambda=810$~nm are produced at angles $\pm 3$~degrees when a photon from a pump laser at 405~nm is incident on a nonlinear BiBO crystal~\cite{Burnham70,Hong86}. The pump laser (200~mW) is rotated to vertical polarization by a half-wave plate (HWP) and incident on the BiBO crystal (0.5~mm thickness). A single photon from the pump produces two `twin' (idler and signal) horizontally polarized photons. One photon is produced in arm A and the other in arm B. The detection of a single photon in arm A using a single-photon detector (Excelitas SPCM-AQRH-15) heralds the presence of a single photon in arm B within an 8~ns coincidence window. A qubit is encoded onto the single photon in arm B using a quarter-wave plate (QWP) and a HWP. Here, the polarization states $\ket{H}$ and $\ket{V}$ are used as the orthogonal basis states of the qubit. This qubit is then sent through the plasmonic metamaterial, after which the state of the qubit is measured via a projective measurement using a QWP, a HWP and a single-photon detector~\cite{James01}. A broadband external control laser (Fianium WhiteLase micro) is used to vary the temperature of the metamaterial by heating it with a range of laser powers.

In order to quantify the performance of the metamaterial as a quantum channel at different temperatures, quantum state tomography is carried out on the output states of four different polarization-encoded qubits $\{ \ket{H}, \ket{V}, \ket{+}=\frac{1}{\sqrt{2}}(\ket{H}+\ket{V}), \ket{R}=\frac{1}{\sqrt{2}}(\ket{H}+i\ket{V}) \}$ sent through, with the density matrices reconstructed via projective measurements~\cite{James01}. The output of a given projective measurement is sent to a single-photon detector and a coincidence between the detector in the heralding arm A and the detector in arm B is measured. Interference filters (810~$\pm$ 5~nm) are placed in front of each detector to cut out photons of higher and lower frequencies corresponding to unwanted down-conversion processes and the pump beam, leading to $\sim 1000$ coincidences per second (for $\ket{H}$ encoded and $\ket{H}$ measured). The density matrices obtained from quantum state tomography of the four probe states are then used to reconstruct the quantum channel of the metamaterial using quantum process tomography~\cite{Chuang97,Nielsen00}, the details of which are discussed later.
\begin{figure}[t]
\centering
\includegraphics[width=8.7cm]{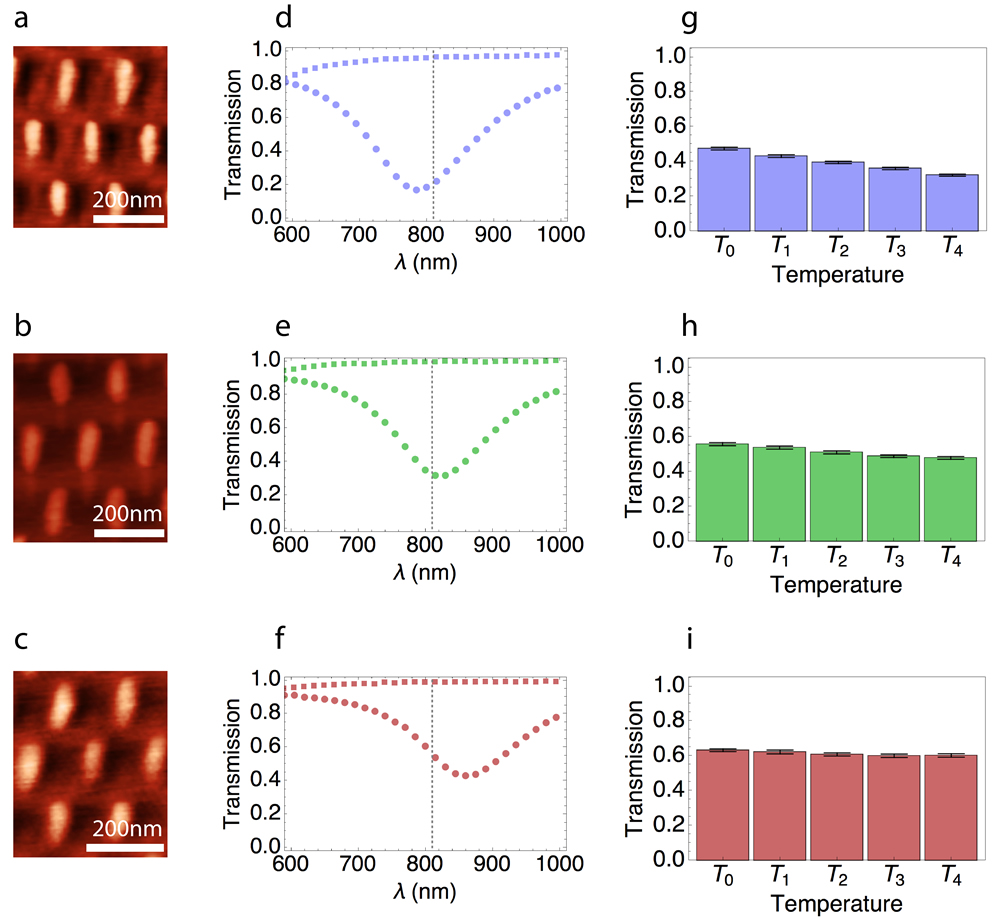}
\caption{\label{fig3} Temperature-dependent transmission response of metamaterials with different nanorod dimensions (experiment). Panels (a), (b) and (c) are atomic force microscope images of three metamaterials used in the experiment. The period and thickness of the nanorods are approximately equal, whereas the width and length increase from (a) to (c). See main text for dimensions. Panels (d), (e) and (f) show classical transmission spectra of the metamaterials at $T_0$ for horizontal (squares) and vertical (circles) polarized light. Panels (g), (h) and (i) are transmission probabilities in the quantum regime for single qubits encoded into the vertical polarization of single photons as $\ket{V}$ and sent through the metamaterials as the temperature is changed. The five different temperature settings are $T_0=295$~K, $T_1=303$~K, $T_2=319$~K, $T_3=331$~K and $T_4=347$~K, corresponding to values consistent with the range used in the theory model. The values are determined by the laser power used and are spaced apart by approximately $10$~K.}
\end{figure}

Three different plasmonic metamaterials were used in this study, each with a specific set of dimensions for the gold nanorod unit cells. The metamaterials were fabricated on an indium tin oxide (ITO)-coated fused silica substrate by electron-beam lithography. A 5~nm thin layer of ITO was deposited on a 5~mm~$\times$~5~mm silica substrate by electron-beam evaporation and then a 200~nm thick film of polymethylmethacrylate photoresist (MicroChem) was spin-coated on top of the ITO. Using electron beam-writing (Raith e-line), the photoresist was patterned and then developed, leaving a mask. Subsequent gold evaporation and lift-off produced the gold nanorod antenna arrays for the different metamaterials, each with an area of 100~$\mu$m~$\times$~100~$\mu$m. The nanorods have a period of 200~nm, a thickness of 30~nm, a width between 50~nm and 70~nm, and a length between 100~nm and 130~nm. Specific dimensions of a given metamaterial are provided later. The full 5~mm~$\times$~5~mm metamaterial sample consisting of an array of metamaterials with different nanorod sizes is placed inside a telescope system designed in such a way that the beam before and after the lenses (planoconvex, $f=25$~mm) is collimated and the beam between the lenses is focused to a spot size with diameter $\lesssim 100~\mu$m.
\begin{figure}[t]
\centering
\includegraphics[width=7.4cm]{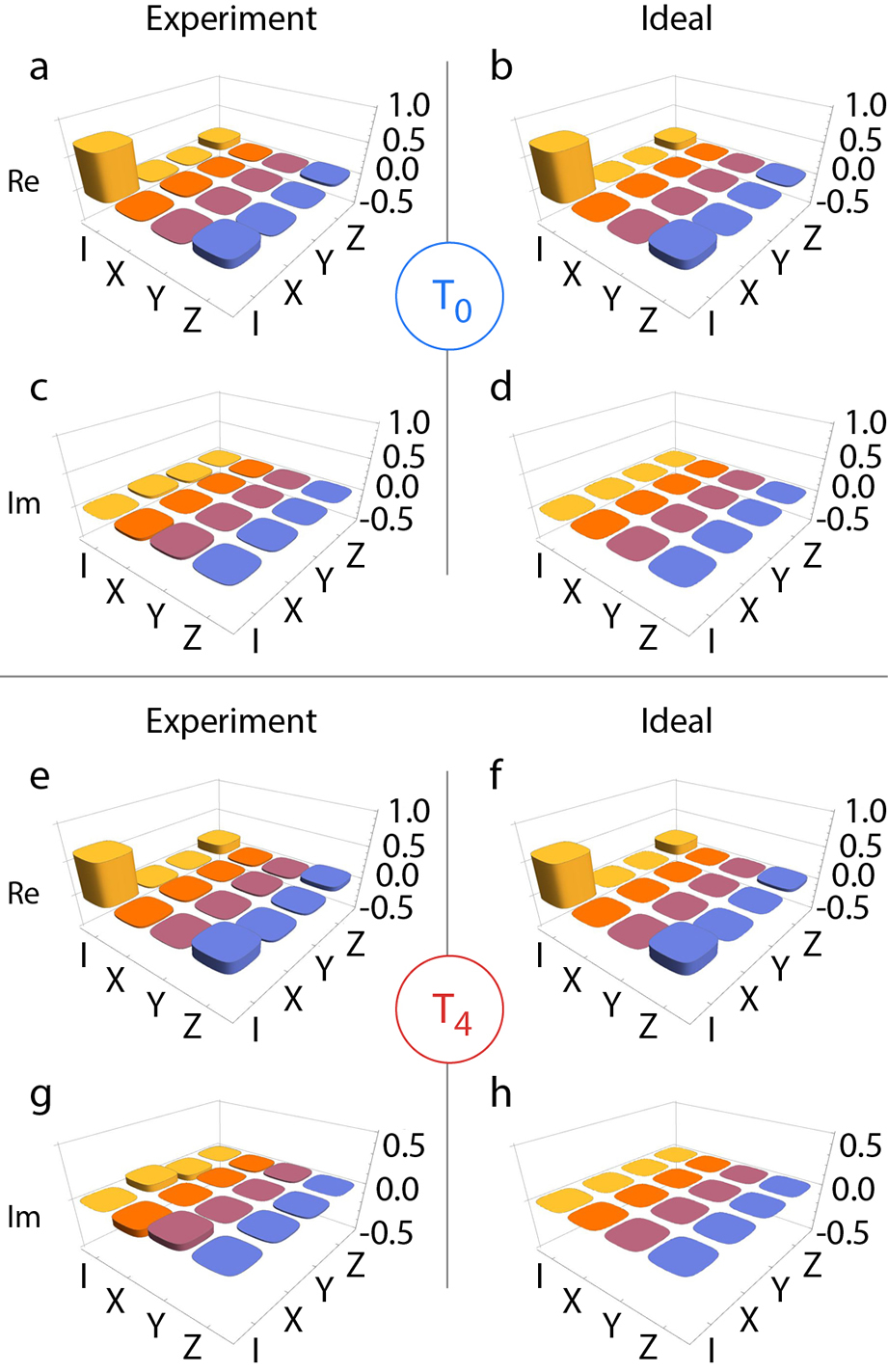}
\caption{\label{fig4} Quantum process tomography $\chi$ matrices for the first metamaterial at two different temperatures, $T_0$ and $T_4$. Panels (a) and (c) show the real and imaginary parts of the experimental $\chi$ matrix at $T_0$. Panels (b) and (d) show the real and imaginary parts of an ideal partial polarizer $\chi$ matrix with $T_V=0.476$. Panels (e) and (g) show the real and imaginary parts of the experimental $\chi$ matrix at $T_4$. Panels (f) and (h) show the real and imaginary parts of an ideal partial polarizer $\chi$ matrix with $T_V=0.324$.}
\end{figure}

{\it Experimental results.---} We now analyse our experimental results in light of the theoretical model described. Here, each of the three metamaterials used has different nanorod dimensions for the unit cells, as shown in the atomic force microscope (AFM) images in Fig.~\ref{fig3}~(a)-(c). The general trend in dimensions is the same as that used in the theoretical model, {\it i.e.} the length and width increase when going from (a) to (c). Due to the background dielectric material not completely encompassing the nanorods, as well as the presence of the ITO bonding layer and differences in the permittivity of gold, it is not possible to obtain an exact fit of our theoretical model to the experimental transmission data. However, the general trend seen in the experimental classical transmission results of Fig.~\ref{fig3}~(d)-(f) matches well that seen in the theoretical model of Fig.~\ref{fig2}~(a)-(c) at $300$~K, also taking into account broadening due to the fabrication process. The dimensions measured by the AFM are $w\simeq 50$~nm and $l\simeq 100$~nm for Fig.~\ref{fig3}~(a), $w\simeq 60$~nm and $l\simeq 110$~nm for (b), and $w\simeq 70$~nm and $l\simeq 130$~nm for (c). The thickness of the nanorods is $30$~nm.

In Fig.~\ref{fig3}~(g), (h) and (i) we show the transmission results of probing the metamaterials with single photons in the state $\ket{V}$ in arm B as the temperature is increased. Here, the transmission is given by the ratio of heralded detection counts (coincidences) when the state $\ket{V}$ is sent through the metasurface and counts when it is sent through the substrate only (no metasurface). It represents the relative probability for a photon in the state $\ket{V}$ to be transmitted through the metamaterial. The temperature is changed by increasing the power of the control laser in 4 steps, from 0~mW to 200~mW, which heats up the metamaterial. The time between the control laser activation and the start of measurements is 480~s for each temperature, however a steady state response is reached within 240~s. For quantum applications such as entanglement distillation~\cite{Asano15,Faro15}, this response time is practical as it is much shorter than the time scale on which birefringent fluctuations would occur in a realistic optical fiber quantum network~\cite{Poppe04}. The response time could be made shorter, if needed for a given application, using alternative heating methods~\cite{Driscoll08,Liu16,Lewan15}. Control via the laser power gives five different temperature settings: $T_0=295$~K, $T_1=303$~K, $T_2=319$~K, $T_3=331$~K and $T_4=347$~K, consistent with the range used in the theoretical model. The values are spaced apart by approximately $10$~K and are determined by the power set by the control laser software. They are measured using a point-probe temperature sensor placed close to the laser beam on the metasurface sample. Measurements are carried out at the steady state response time and the error in the values is $<1$~K, consistent with ambient temperature fluctutations. 
 \begin{table*}[t]
  \centering
    \begin{tabular}{|c|c|c|c|c|c|c|c|c|c|c|}
    \hline
          & \multicolumn{3}{c|}{$M_1$} & \multicolumn{3}{c|}{$M_2$} & \multicolumn{3}{c|}{$M_3$} \\
          \hline
   Temp & $F_P$    & $T_H$ & $T_V$   & $F_P$    & $T_H$ & $T_V$    & $F_P$    & $T_H$ & $T_V$ \\
    \hline
    $T_0$     & 0.935 $\pm$ 0.007 & 1     & 0.476 $\pm$ 0.008 & 0.909 $\pm$ 0.012 & 1     & 0.560 $\pm$ 0.009 & 0.895 $\pm$ 0.009 & 1     & 0.634 $\pm$ 0.008 \\
    \hline
    $T_1$   & 0.899 $\pm$ 0.006 & 1     & 0.433 $\pm$ 0.008 & 0.935 $\pm$ 0.006 & 1     & 0.540 $\pm$ 0.009 & 0.922 $\pm$ 0.007 & 1     & 0.624 $\pm$ 0.011 \\
    \hline
    $T_2$   & 0.911 $\pm$ 0.006 & 1     & 0.397 $\pm$ 0.007 & 0.934 $\pm$ 0.008 & 1     & 0.513 $\pm$ 0.009 & 0.942 $\pm$ 0.009 & 1     & 0.610 $\pm$ 0.009 \\
    \hline
    $T_3$  & 0.922 $\pm$ 0.005 & 1     & 0.362 $\pm$ 0.007 & 0.948 $\pm$ 0.008 & 1     & 0.491 $\pm$ 0.008 & 0.970 $\pm$ 0.006 & 1     & 0.602 $\pm$ 0.010 \\
    \hline
    $T_4$  & 0.897 $\pm$ 0.005 & 1     & 0.324 $\pm$ 0.006 & 0.912 $\pm$ 0.006 & 1     & 0.481 $\pm$ 0.008 & 0.939  $\pm$ 0.009 & 1     & 0.604 $\pm$ 0.010 \\
    \hline
    \end{tabular}
  \caption{Process fidelities for the three metamaterials investigated as the temperature is changed, as well as horizontal and vertical transmission probabilities $T_H$ and $T_V$ extracted from maximizing the process fidelity, respectively.}
    \label{table:tab1}
\end{table*}

At $T=T_0$, one can see in Fig.~\ref{fig3}~(g), (h) and (i) that the photon transmission slightly deviates from that of the classical transmission measured using vertically polarized light at $\lambda=810$~nm, as shown in Fig.~\ref{fig3}~(d), (e) and (f). This deviation can be attributed to the spot size of the beam -- in the classical case the spot is smaller and easier to align on the metasurface using a CCD camera, whereas in the single-photon case the spot size is comparable to the metasurface and alignment is achieved by optimising single-photon detection counts. As a result there is some non-ideal overlap of the beam and the metamaterial. Regardless of this, the trend of the single-photon transmission at $T_0$ matches that of the classical case as the nanorod dimensions increase. Moreover, as the temperature increases one can see the effect on the transmission of $\ket{V}$ states for the three metamaterials considered. The largest change is seen for the first metamaterial, shown in Fig.~\ref{fig3}~(g), where the transmission changes from $0.48$ to $0.32$, corresponding to a percentage change of $33~\%$. The percentage changes for the other two metamaterials are $14\%$ and $5 \%$. We also measured the transmission of $\ket{H}$ states through the metamaterials as the temperature was changed and found that the transmission remained roughly the same as when the states were sent through the substrate only. The exact transmission values of the $\ket{H}$ state, as well as those of the additional probe states $\ket{D}$ and $\ket{R}$ are combined with the values obtained for the $\ket{V}$ state to obtain a full characterization of the metamaterial as a variable single-qubit quantum channel. The transmission values are part of a larger set of projective measurements which we use for quantum process tomography~\cite{Chuang97,Nielsen00} at the five different temperatures and discussed in detail next.

The four probe states sent through the metamaterial in the quantum process tomography are $\ket{H}$, $\ket{V}$, $\ket{D}$ and $\ket{R}$. From projective measurements on the outputs of these states in the bases $\ket{H/V}$, $\ket{D/A}$ and $\ket{R/L}$, we reconstructed their density matrices using quantum state tomography~\cite{James01}. Using the density matrices we then obtained the quantum process matrices, or $\chi$ matrices, for the three different metamaterials in our investigation~\cite{Chuang97}. The polarization response of the metamaterials is such that they act as partial polarizers and are well represented by a single Kraus operator $K_0$ = $\ket{H}\bra{H}$ + $\sqrt{T_V}\ket{V}\bra{V}$ corresponding to a non-trace preserving channel~\cite{Asano15}, {\it i.e.} $\rho \rightarrow K_0 \rho K_0^ \dagger$, where $\rho$ is the qubit of the input single-photon state in the polarization basis. This channel is equivalent to the general quantum channel $\rho \rightarrow \sum_{ij}\chi_{ij} E_i \rho E_j^ \dagger$, where the single-qubit Pauli operators, $E_i = I, X, Y$ and $Z$, provide a complete basis for the Hilbert space and the elements of the $\chi$ matrix are set to values that allow the general channel to completely match the action of $K_0$~\cite{Asano15}. 

The $\chi$ matrix obtained for the first metamaterial at $T_0$ is shown in Fig.~\ref{fig4}~(a) and (c). Fig.~\ref{fig4}~(a) shows the real part and (c) shows the imaginary part. The real and imaginary parts of an ideal partial polarizer matrix $\chi_{id}$ with $T_V=0.476$ are shown in Fig.~\ref{fig4}~(b) and (d), respectively. The value of $T_V$ has been found using the process fidelity, $F_P(T_V)= {\rm Tr} (\sqrt{\sqrt{\chi}\chi_{id}\sqrt{\chi}})^2 / {\rm Tr}(\chi){\rm Tr}(\chi_{id})$, which quantifies how close the experimental channel is to an ideal channel of a partial polarizer. We find a maximum of $F_P(T_V)=0.935 \pm 0.008$ at $T_V=0.476\pm 0.008$, which shows that the metamaterial represents well a partial polarizer for single photons with a $T_V$ value consistent with the single-photon transmission measured previously (see Fig.~\ref{fig3}~(g)). The $\chi$ matrix for the first metamaterial at $T_4$ is shown in Fig.~\ref{fig4}~(e) and (g). Fig.~\ref{fig4}~(e) shows the real part and (g) shows the imaginary part. The real and imaginary parts of an ideal partial polarizer matrix with $T_V=0.324$ are shown in Fig.~\ref{fig4}~(b) and (d). The value of $T_V$ has again been found by maximising the process fidelity, with a value of $F_P=0.897 \pm 0.005$. The process fidelities and corresponding $T_V$ values extracted for all three metamaterials at all temperatures are given in Tab.~\ref{table:tab1}. All process fidelities are above $89 \%$, with $T_V$ consistent with the values measured previously (see Fig.~\ref{fig3}~(g), (h) and (i)), showing the metamaterials act as variable partial polarizers in the quantum regime. As a result, they can be used to induce a temperature-controlled collective polarization-dependent loss at the single-photon level for quantum information tasks, such as entanglement distillation~\cite{Asano15}.
  
{\it Summary.---} We investigated the active control of a plasmonic metamaterial in the quantum regime via its thermal response. Metamaterials with unit cells made from gold nanorods were probed with qubits encoded into single photons. Using an external laser we controlled the temperature of the nanorods and substrate. We then carried out quantum process tomography, characterizing the metamaterials as variable quantum channels. It was found that the overall polarization response of the metamaterials can be tuned by up to 33\% for particular nanorod dimensions. We used a theoretical model to describe the thermal response of the metamaterials and found that our experimental results matched the predicted behavior well. Our work goes beyond previous studies of simple passive plasmonic systems in the quantum regime and shows that external control of plasmonic elements provides variable metamaterials that can be used for quantum state engineering tasks.

{\it Acknowledgments.---} This research was supported by the South African National Research Foundation, the National Laser Centre, the UKZN Nanotechnology Platform, the South African National Institute for Theoretical Physics, MEXT/JSPS KAKENHI Grant Number No. JP15KK0164 and No. JP18H04291, the Helmholtz program Science and Technology of Nanosystems (STN) and the Karlsruhe School of Optics \& Photonics (KSOP). S.~K.~O is supported by ARO grant No. W911NF-18-1-0043 and Pennsylvania State University Materials Research Institute (MRI).



\begin{thebibliography}{99}

\bibitem{Lee12} J. H. Lee, J. P. Singer and E. L. Thomas, Micro-nanostructured mechanical metamaterials, Adv. Mater. {\bf 24}, 4782-4810 (2012).

\bibitem{Christ15} J. Christensen, M. Kadic, M. Wegener and O. Kraft, Vibrant times for mechanical metamaterials, MRS Commun. {\bf 5}, 453-462 (2015).

\bibitem{Cummer16} S. A. Cummer, J. Christensen and A. Al\`u, Controlling sound with acoustic metamaterials, Nature Rev. Mater. {\bf 1}, 16001 (2016).

\bibitem{Cai10} W. Cai and V. Shalaev, Optical Metamaterials: Fundamentals and applications (Springer, Dordrecht, 2010).

\bibitem{Soukoulis11} C. M. Soukoulis and M. Wegener, Past achievements and future challenges in the development of three-dimensional photonic metamaterials, Nature Photonics {\bf 5}, 523 (2011).

\bibitem{Maier07} S. A. Maier, Plasmonics: Fundamentals and Applications, (Springer, New York, 2007).

\bibitem{Jahani16} S. Jahani and Z. Jacob, All-dielectric metamaterials, Nature Nanotech. {\bf 11}, 23-36 (2016).

\bibitem{Zhang08} X. Zhang and Z. Liu, Superlenses to overcome the diffraction limit, Nature Mater. {\bf 7}, 435-441 (2008).

\bibitem{Chen10} H. Chen, C. T. Chan and P. Sheng, Transformation optics and metamaterials, Nature Mater. {\bf 9}, 387 (2010).

\bibitem{Chen12} T. Chen, S. Li and H. Sun, Metamaterials Application in Sensing, Sensors {\bf 12}, 2742-2765 (2012).

\bibitem{Billings13} L. Billings, Metamaterial world, Nature {\bf 500}, 138 (2013).

\bibitem{Wang12} S. M. Wang, S. Y. Mu, C. Zhu, Y. X. Gong, P. Xu, H. Liu, T. Li, S. N. Zhu and X. Zhang, Hong-Ou-Mandel interference mediated by the magnetic plasmon waves in a three-dimensional optical metamaterial. Opt. Exp. {\bf 20}, 5213 (2012).

\bibitem{Zhou12} Z.-Y. Zhou, D.-S. Ding, B.-S. Shi, X.-B. Zou and G. C. Guo, Characterizing dispersion and absorption parameters of metamaterial using entangled photons, Phys. Rev. A {\bf 85}, 023841 (2012).

\bibitem{Chipouline12} A. Chipouline, S. Sugavanam, V. A. Fedotov and A. E. Nikolaenko, Analytical model for active metamaterials with quantum ingredients, J. Opt. {\bf 14}, 114005 (2012).

\bibitem{Cortes14} C. L. Cortes, W. Newman, S. Molesky and Z. Jacob, Quantum nanophotonics using hyperbolic metamaterials, J. Opt. {\bf 16}, 129501 (2014).

\bibitem{Scholl15} J. A. Scholl, A. Garcia-Etxarri, G. Aguirregabiria, R. Esteban, T. C. Narayan, A. L. Koh, J. Aizpurua and J. A. Dionne, Evolution of Plasmonic Metamolecule Modes in the Quantum Tunneling Regime, ACS Nano {\bf 10}, 1346-1354 (2016).

\bibitem{McEnery14} K. McEnery, M. S. Tame, S. A. Maier and M. S. Kim, Tunable negative permeability in a quantum plasmonic metamaterial, Phys. Rev. A {\bf 89}, 013822 (2014).

\bibitem{Moiseev10} S. A. Moiseev, A. Kamli and B. C. Sanders, Low-loss nonlinear polaritonics, Phys. Rev. A. {\bf 81}, 033839 (2010).

\bibitem{Siomau12} M. Siomau, A. Kamli, S. A. Moiseev and B. C. Sanders, Entanglement creation with negative index metamaterials, Phys. Rev. A {\bf 85}, 050303 (2012).

\bibitem{Zhou16} M. Zhou, J. Liu, M. Kats and Z. Yu, Atomic metasurfaces for manipulation of single photons, ACS Photonics {\bf 4}, 1279 (2017).

\bibitem{Tame13} M. S. Tame, K. R. McEnery, \c{S}. K. \"{O}zdemir, S. A. Maier and M. S. Kim, Quantum Plasmonics, Nature Physics {\bf 9}, 329 (2013).

\bibitem{Zago16} A. M. Zagoskin, D. Felbacq and E. Rousseau, Quantum metamaterials in the microwave and optical ranges, EPJ Quant. Tech. {\bf 3}, 2 (2016).

\bibitem{Asano15} M. Asano, M. Bechu, M. Tame, S. K. \"Ozdemir, R. Ikuta, D. \"O. G\"uney, T. Yamamoto, L. Yang, M. Wegener and N. Imoto, Distillation of photon entanglement using a plasmonic metamaterial, Sci. Rep. {\bf 5}, 18313 (2015).

\bibitem{Faro15} Md Abdullah al Farooqui, J. Breeland, M. I. Aslam, M. Sadatgol, \c{S}. K. \"Ozdemir, M. S. Tame, L. Yang and D. \"O. G\"uney, Quantum entanglement distillation with metamaterials, Opt. Exp. {\bf 23}, 17941-17954 (2015).

\bibitem{Jha15} P. K. Jha, X. Ni, C. Wu, Y. Wang and X. Zhang, Metasurface enabled remote quantum interference, Phys. Rev. Lett. {\bf 115}, 025501 (2015).

\bibitem{Gheer16} N. Gheeraert, S. Bera and S. Florens, Spontaneous emission of many-body Schrödinger cats in metamaterials with large fine structure constant, New J. Phys. {\bf 19}, 023036 (2017).

\bibitem{Podd16} A. N. Poddubny, I. V. Iorsh and A. A. Sukhorukov, Generation of photon-plasmon quantum states in nonlinear hyperbolic metamaterials, Phys. Rev. Lett. {\bf 117}, 123901 (2016).

\bibitem{Roger15} T. Roger, S. Vezzoli, E. Bolduc, J. Valente, J. J. F. Heitz, J. Jeffers, C. Soci, J. Leach, C. Couteau, N. I. Zheludev and D. Faccio, Coherent perfect absorption in deeply subwavelength films in the single-photon, Nature Commun. {\bf 6}, 7031 (2015).

\bibitem{Altuzarra17} C. Altuzarra, S. Vezzoli, J. Valente, W. Gao, C. Soci, D. Faccio and C. Couteau, Nonlocal control of dissipation with entangled photons, arXiv:1701.05357 (2017).

\bibitem{Boardman11} A. D. Boardman, V. V. Grimalsky, Y. S. Kivshar, S. V. Koshevaya, M. Lapine, N. M. Litchinitser, V. N. Malnev, M. Noginov, Y. G. Rapoport and V. M. Shalaev, Active and tunable metamaterials, Las. Phot. Rev. {\bf 5}, 287-307 (2011).

\bibitem{Hess12} O. Hess, J. B. Pendry, S. A. Maier, R. F. Oulton, J. M. Hamm and K. L. Tsakmakidis, Active nanoplasmonic metamaterials, Nature Mat. {\bf 11}, 573-584 (2012).

\bibitem{Zheludev12} N. I. Zheludev and Y. S. Kivshar, From metamaterials to metadevices, Nature Mat. {\bf 11}, 917-924 (2012).

\bibitem{Gavinit72} A. Gavinit and C. C. Y. Kwan, Optical Properties of Semiconducting VO2 Films, Phys. Rev. B {\bf 5}, 3138 (1972).

\bibitem{Jepsen06} P. Uhd Jepsen, B. M. Fischer, A. Thoman, H. Helm, J. Y. Suh, R. Lopez and R. F. Haglund, Jr., Metal-insulator phase transition in a VO2 thin film observed with terahertz spectroscopy, Phys. Rev. B {\bf 74}, 205103 (2006).

\bibitem{Prayakarao16} S. Prayakarao, B. Mendoza, A. Devine, C. Kyaw, R. B. van Dover, V. Liberman and M. A. Noginov, Tunable VO2/Au hyperbolic metamaterial, App. Phys. Lett. {\bf 109}, 061105 (2016).

\bibitem{Wang15} Q. Wang, E. T. F. Rogers, B. Gholipour, C.-M. Wang, G. Yuan, J. Teng and N. I. Zheludev, Optically reconfigurable metasurfaces and photonic devices based on phase change materials, Nature Photon. {\bf 10}, 60 (2015).

\bibitem{Werner07} D. H. Werner, D.-H. Kwon and I.-C. Khoo, A. V. Kildishev and V. M. Shalaev, Liquid crystal clad near-infrared metamaterials with tunable negative-zero-positive refractive indices, Opt. Exp. {\bf 15}, 3342 (2007).

\bibitem{Khoo10} I. C. Khoo, A. Diaz, J. Liou, M. V. Stinger, J. Huang and Y. Ma, Liquid Crystals Tunable Optical Metamaterials, IEEE J. Sel. Top. Quant. Elec. {\bf 16}, 410 (2010).

\bibitem{Shrek13} D. Shrekenhamer, W.-C. Chen and W. J. Padilla, Liquid Crystal Tunable Metamaterial Absorber, Phys. Rev. Lett. {\bf 110}, 177403 (2013).

\bibitem{Gutruf16} P. Gutruf, C. Zou, W. Withayachumnankul, M. Bhaskaran, S. Sriram and C. Fumeaux, Mechanically Tunable Dielectric Resonator Metasurfaces at Visible Frequencies, ACS Nano {\bf 10}, 133 (2016).

\bibitem{Ee16} H.-S. Ee and R. Agarwal, Tunable Metasurface and Flat Optical Zoom Lens on a Stretchable Substrate, Nano Lett. {\bf 16}, 2818-2823 (2016).

\bibitem{Liu17} W. Liu, Y. Shen, G. Xiao, X. She, J. Wang and C. Jin, Mechanically tunable sub-10nm metal gap by stretching PDMS substrate, Nanotechnology {\bf 28}, 075301 (2017).

\bibitem{Kamali16} S. M. Kamali, E. Arbabi, A. Arbabi, Y. Horie and A. Faraon, Highly tunable elastic dielectric metasurface lenses, Laser Photon. Rev. {\bf 10}, 1002-1008 (2016).

\bibitem{Feig10} E. Feigenbaum, K. Diest and H. A. Atwater, Unity-Order Index Change in Transparent Conducting Oxides at Visible Frequencies, Nano Lett. {\bf 10}, 2111-2116 (2010).

\bibitem{Poppe04} A. Poppe, A. Fedrizzi, T. Loruenser, O. Maurhardt, R. Ursin, H. R. Boehm, M. Peev, M. Suda, C. Kurtsiefer, H. Weinfurter, T. Jennewein and A. Zeilinger, Practical Quantum Key Distribution with Polarization-Entangled Photons, Opt. Express {\bf 12}, 3865-3871 (2004).

\bibitem{Driscoll08} T. Driscoll, S. Palit, M. M. Qazilbash, M. Brehm, F. Keilmann, B.-G. Chae, S.-J. Yun, H.-T. Kim, S. Y. Cho, N. M. Jokerst, D. R. Smith and D. N. Basov, Dynamic tuning of an infrared hybrid-metamaterial resonance using vanadium dioxide, App. Phys. Lett. {\bf 93}, 024101 (2008).

\bibitem{Liu16} X. Liu and W. J. Padilla, Thermochromic Infrared Metamaterials, Adv. Mater. {\bf 28}, 871 (2016).

\bibitem{Lewan15} W. Lewandowski, M. Fruhnert, J. Mieczkowski, C. Rockstuhl and E. G\'orecka1, Dynamically self-assembled silver nanoparticles as a thermally tunable metamaterial, Nat. Comm. {\bf 6}, 6590 (2015).

\bibitem{Kildishev13} A. V. Kildishev, A. Boltasseva and V. M. Shalaev, Planar Photonics with Metasurfaces, Science {\bf 339}, 1232009 (2013).

\bibitem{Yu14} N. Yu and F. Capasso, Flat optics with designer metasurfaces, Nature Mater. {\bf 13}, 139 (2014).

\bibitem{Meinzer14} N. Meinzer, W. L. Barnes and I. R. Hooper, Plasmonic meta-atoms and metasurfaces, Nature Photon. {\bf 8}, 889 (2014).

\bibitem{Xia14} F. Xia, H. Wang, D. Xiao, M. Dubey and A. Ramasubramaniam, Two-dimensional material nanophotonics, Nature Photon. {\bf 8}, 899 (2014).

\bibitem{Lin14} D. Lin, P. Fan, E. Hasman and M. L. Brongersma, Dielectric gradient metasurface optical elements, Science {\bf 345}, 298 (2014).

\bibitem{Gisin07} N. Gisin and R. Thew, Quantum communication, Nature Photon. {\bf 1}, 165-171 (2007).

\bibitem{Burnham70} D. C. Burnham and D. L. Weinberg, Observation of simultaneity in parametric production of optical photon pairs, Phys. Rev. Lett {\bf 25}, 84-87 (1970).

\bibitem{Hong86} C. K. Hong and L. Mandel, Experimental realization of a localized one-photon state, Phys. Rev. Lett {\bf 56}, 58-60 (1986).

\bibitem{Alu11} A. Al\`u and N. Engheta, Ch. 3 in Structured surfaces as optical metamaterials, Ed. A. A. Maradudin, Cambridge University Press (2011).

\bibitem{Zhao11} Y. Zhao, N. Engheta and A. Al\`u, Homogenization of plasmonic metasurfaces modeled as transmission-line loads, Metamaterials {\bf 5}, 90 (2011).

\bibitem{Zhao13} Y. Zhao and A. Al\`u, Tailoring the Dispersion of Plasmonic Nanorods To Realize Broadband Optical Meta-Waveplates, Nano Lett. {\bf 13}, 1086 (2013).

\bibitem{Bohren83} C. F. Bohren and D. R. Huffman, Absorption and scattering of light by small particles, Wiley (1983).

\bibitem{Lin07} K.-Q. Lin, L.-M. Wei, D.-G. Zhang, R.-S. Zheng, P. Wang, Y.-H. Lu and H. Ming, Temperature effects on prism-nased surface plasmon resonance sensor, Chin. Phys. Lett. {\bf 24}, 3081 (2007).

\bibitem{Leviton08} D. B. Leviton and B. J. Frey, Temperature-dependent absolute refractive index measurements of synthetic fused silica, Proc. SPIE 6273 June (2008).

\bibitem{Rakic98} A. D. Raki\'c, A. B. Djuri{\v s}i\'c, J. M. Elazar and M. L. Majewski, Optical properties of metallic films for vertical-cavity optoelectronic devices, Appl. Opt. {\bf 37}, 5271 (1998).

\bibitem{Ozdemir03} \c{S}. K. \"Ozdemir and G. Turhan-Sayan, Temperature Effects on Surface Plasmon Resonance: Design Considerations for an Optical Temperature Sensor, J. Light. Tech. {\bf 21}, 805 (2003).

\bibitem{Bouillard12} J.-S. G. Bouillard, W. Dickson, D. P. O'Connor, G. A. Wurtz and A. V. Zayats, Low-Temperature Plasmonics of Metallic Nanostructures, Nano Lett. {\bf 12}, 1561 (2012).

\bibitem{James01} D. F. V. James, P. G. Kwiat, W. J. Munro, and A. G. White. Measurement of qubits. Phys. Rev. A {\bf 64}, 052312 (2001).

\bibitem{Chuang97} I. L. Chuang and M. A. Nielsen. Prescription for experimental determination of the dynamics of a quantum black box. J. Mod. Opt. {\bf 44}, 2455 (1997).

\bibitem{Nielsen00} M. A. Nielsen and I. L. Chuang, {\sl Quantum Computation and Quantum Information}, Cambridge University Press, Cambridge (2000).

\end{thebibliography}
\end{document}